# THE AGES OF STARBURSTS IN POST-STARBURST GALAXIES

Andrew J. Leonardi

Department of Physics and Astronomy, University of North Carolina, Chapel Hill, NC 27599

James A. Rose[1]

Department of Physics and Astronomy, University of North Carolina, Chapel Hill, NC 27599

## ABSTRACT

We present a new technique for accurately determining the ages of starbursts in post-starburst galaxies. In particular, it uses the strength of the Ca II H + H$\epsilon$ absorption feature relative to that of Ca II K to separate the effects of burst strength from burst age in a spectrum comprised of a post-starburst population and an underlying old galaxy population. The technique is based on comparing the integrated spectra of post-starburst galaxies with starburst model spectra produced from the evolutionary synthesis code of Bruzual & Charlot (1995, ApJ, in preparation). Model spectra have been generated for starbursts viewed at a variety of times (ranging from 0.0 to 2.0 Gyr) after completion of the burst. Each of these spectra has been combined in various amounts with observed integrated spectra of elliptical galaxies to simulate different combinations of starbursts and underlying old populations.

To test the technique we have measured spectral indices in the observed spectra of three post-starburst early-type galaxies. The three galaxies are shown to have different post-starburst ages, ranging from 0 Gyr to 1.5 Gyr after completion of the burst.

We also demonstrate that if the burst contributes greater than 50% of the light at 4000 Å then the technique can distinguish between a post-starburst systems and models in which a constant star formation history is suddenly truncated, such as that which could occur in a spiral that is rapidly stripped of its interstellar medium.

Finally, we demonstrate that while the technique is well-suited for determining the time elapsed since the termination of a starburst (up until 2 Gyr after the end of the burst), the derived termination times are relatively insensitive to the duration of the burst.

## 1. Introduction

Abnormally powerful star formation events (hereafter referred to as starbursts) have been suspected as the underlying cause for the unusually blue colors observed in some galaxies. Since

astro-ph/9510033  05 Oct 1995

Searle, Sargent & Bagnuolo (1973) proposed the starburst idea to explain the bluest dwarf galaxies known at that time, the starburst concept has gained in popularity and is now used to denote any vigorous episode of star formation caused by a distinct triggering event and lasting for only a short period of time. While much attention concerning starburst galaxies has been focussed on interacting and merger galaxies at low redshift, it has also been known since the seminal studies of Butcher & Oemler (1978, 1984) that distant, rich galaxy clusters have a much higher fraction of photometrically blue galaxies than present-day clusters. Spectroscopic studies of clusters at redshift z∼0.3-0.5 have now demonstrated that many of the blue galaxies exhibit enhanced Balmer absorption lines but with weak or absent emission lines, thereby distinguishing them from the spectra of ordinary spiral galaxies (e.g., Dressler and Gunn 1983; Lavery and Henry 1986; Couch and Sharples 1987). The term "E+A" has been coined to refer to these galaxies because spectroscopically they appear to have an A-star component superimposed on an old, elliptical galaxy spectrum (e.g., Gunn and Dressler 1988). Furthermore, it is proposed that the E+A galaxies underwent an intense period of star formation ∼1-2 Gyr prior to the epoch of observation.

Recently, E+A galaxies have also been found in the nearby Coma cluster, during a survey of early-type galaxies (Caldwell et al. 1993). In the Coma cluster, the majority of the galaxies with E+A spectra similar to those observed by Dressler & Gunn (1983) are located in the southwest region of the cluster between the main cluster center and the secondary X-ray peak discovered by Briel et al. (1992) 40' to the southwest. On the basis of N-body and gas dynamical simulations, Burns et al. (1994) have proposed that the E+A galaxies belong to a substructure that has passed through the cluster center and emerged out the other side. They suggest the group is now falling back in again. The time scale for this event to occur according to the simulations is about 2 Gyr. Clearly, information concerning the ages of the post-starburst events can serve as a useful clock to compare with the ∼2 Gyr age derived from dynamical considerations. To make such a comparison between dynamical and star-formation timescales, an accurate method for determining the ages of starbursts is required.

In practice, it has proven difficult to reliably determine the ages and burst strengths in post-starburst galaxies (e.g., Couch and Sharples 1987; Bica et al. 1990a; Newberry et al. 1990; Schweizer and Seitzer 1992; Charlot and Silk 1994). The most serious problem involves differentiating between weak, young starbursts and old, strong starbursts, both of which can produce similar broadband colors and Balmer line strengths. In effect, there exists a degeneracy between the competing effects of age and strength of the starburst that is reminiscent of the age-metallicity degeneracy that has plagued studies of old stellar populations in integrated light, such as elliptical galaxies (Worthey 1994 and references therein). Another difficulty in using broadband colors to determine the age of starbursts is that they so rapidly return to their pre-burst values. Specifically, Bica et al. (1990a) and Charlot & Silk (1994) have shown that colors become virtually indistinguishable from their pre-burst values within 1 Gyr for a burst mass of up to 10% of the total mass of the galaxy. Thus to date it has been difficult to assess with confidence the ages of post-starburst galaxies. Studies such as those of Bica et al. (1990a) and Charlot & Silk (1994) have emphasized the necessity of using individual absorption features in addition to the broadband colors.

In this paper a method is presented for determining the epoch of starbursts in E+A galaxies. It is then applied to the E+A galaxies in the Coma cluster in a companion paper (Caldwell et al. 1995). Our technique uses galaxy evolution models to trace the evolution of two spectral indices that, taken together, unambiguously quantify the presence of a young stellar component in the integrated spectrum of a galaxy. By measuring these indices in post-starburst galaxies we are able to determine both the time elapsed since the end of the starburst and the strength of the burst in a manner that avoids the degeneracy between age and burst strength mentioned above. In §2 we describe the spectral indices used and the procedures used to model them in post-starburst galaxies. In §3 we summarize observations of three E+A galaxies, and in §4 we apply the age-dating method to those observations to test its utility. A brief summary is given in §5.

## 2. Age-Dating Technique

### 2.1. Spectral Indices

The age-dating technique that is presented in this paper is based on two of the spectral indices developed and calibrated in Rose (1984, 1985). Each of these indices is defined by taking the ratio of the counts in the bottoms of two neighboring lines *without* reference to the continuum levels. Because only neighboring spectral absorption features are compared, the indices are virtually insensitive to reddening. The first spectral index, hereafter referred to as the H$\delta$/$\lambda$4045 index, is essentially a measure of the spectral type of a star (or of the integrated spectral type of a galaxy); it is produced from the ratio of the central intensity in H$\delta$ relative to that of the neighboring Fe I $\lambda$4045 line. Due to the manner in which the index is defined, it *decreases* in value as one proceeds from late type stars to earlier type stars, i.e., as H$\delta$ strengthens and Fe I $\lambda$4045 weakens. The H$\delta$/$\lambda$4045 index decreases smoothly from spectral type K through A0, where it reaches a minimum value. For spectral types earlier than A0, H$\delta$/$\lambda$4045 increases again, due to the fact that the Balmer lines weaken in B and O stars.

The second index, which is formed from the ratio of the Ca II H + H$\epsilon$ line relative to Ca II K, exhibits a rather different behavior from the H$\delta$/$\lambda$4045 index. This index was originally referred to as the Ca II H + H$\epsilon$/Ca II K index by Rose (1984) but is hereafter referred to as Ca II. As shown by Rose (1985), it is constant in stars later than about F2, but then decreases dramatically for earlier type stars as the Ca II lines weaken and H$\epsilon$ strengthens. The index also reaches a minimum at spectral type A0 and then increases towards earlier spectral types as H$\epsilon$ fades at hotter temperatures. Due to the constant value of the Ca II index in F-K stars it provides an unambiguous signature for stars hotter than F2 in the integrated light of a stellar system if the index falls below the constant value for cool stars. If ongoing star formation or nuclear activity causes emission line filling of H$\epsilon$ however, the index no longer follows the pattern given above and cannot reliably be used in the age-dating method. Rose (1985) used this index to constrain the contribution of hot stars to the integrated light of elliptical galaxies. Recently, Charlot & Silk (1994) have simulated the evolutionary behavior of the index in post-starburst galaxies using the Bruzual & Charlot (1993) evolutionary synthesis models as an example of an absorption line index. They demonstrated that for both a model elliptical galaxy and a model spiral galaxy which undergo a starburst involving 10% of the mass of the galaxy, the U - V and V - K colors approach the values in quiescent galaxies within about 0.5 Gyr after the burst ceases while the Ca II index continues to evolve significantly for 1.5-2 Gyr. Alone, the Ca II index cannot remove the age-strength degeneracy. A second spectral index like the H$\delta/\lambda 4045$ index is needed to decouple the age and strength of the starburst. Below we present models which track the evolution of the Ca II and H$\delta/\lambda 4045$ indices in post-starburst systems.

## 2.2. Starburst Models

The indices are computed from synthetic spectra that are created using an evolutionary synthesis code (Bruzual and Charlot 1995) kindly provided by G. Bruzual. Each synthetic spectrum represents a different starburst scenario characterized by the length of the burst and the time since the burst ended. The spectral library used to compute the synthetic spectra consists of members of the Jacoby, Hunter, & Christian (1984) stellar library.

During the model starburst event, the SFR remains constant, dropping to zero at burst end. A six segment IMF devised by Scalo (1986) was employed. For stars formed in the burst, the lower and upper limits of stellar mass are 0.10 M$_\odot$ and 125 M$_\odot$, respectively. A spectrum was produced for each burst length at 0.0, 0.5, 1.0, 1.5, and 2.0 Gyr after the burst ceased. Each model spectrum was interpolated to produce a 1 Å per pixel dispersion and normalized at 4000 Å. They were then smoothed to a final resolution of 7 Å to match the galaxy observations described in §3. The primary burst length investigated was 0.3 Gyr. This is the typical length found by Mihos et al. (1992, 1993) in their simulations of starbursts induced by galaxy interactions/mergers. However, as will be shown later, the results we obtain are nearly insensitive to the length of the starburst.

By using the galaxy evolution models to produce an integrated galaxy spectrum, this age-dating technique attains a higher level of accuracy than in the use of a single stellar spectrum, such as an A star, to represent the starburst. To illustrate this fact, in Fig. 1 we compare a synthetic spectrum of a 0.3 Gyr starburst model with an A5 stellar spectrum (HR6570, that was kindly provided by L. Jones). Fig. 1b shows a 0.3 Gyr starburst viewed 1.0 Gyr after the burst terminated. The G-band and other metallic features labelled in the figure are prominent in the case of the burst seen 1 Gyr after its termination. In contrast, in the A star spectrum (Fig. 1a) these metallic features are drastically reduced (note the significant Ca II K line though). Usually, a starburst will be seen superimposed on an underlying old population in a galaxy. Hence if an E+A galaxy were modelled with an A star spectrum for the burst and a K giant spectrum for the old population, the result would be to somewhat overestimate the contribution of the old population due to the relative weakness of metallic features in the A star spectrum compared to a concentrated burst of star formation.

Finally, it should be emphasized at this point that the Bruzual & Charlot (1995) models are for solar abundance chemical composition only. Thus we have been constrained to modelling solar abundance populations and have not been able to directly estimate the effect that metal abundance may have on our results. We return to this point in §4.

## 3. Observational Data

To provide a benchmark for comparison with the model starburst spectra, we have obtained integrated spectra in the blue of three early-type galaxies with strong post-starburst spectral signatures. First, an integrated spectrum of the dwarf elliptical galaxy NGC 205 was obtained with the University of Hawaii 2.2-meter telescope at the Mauna Kea Observatory in 1984. This spectrum, which was recorded photographically behind a Carnegie image-tube, covers the wavelength region $\lambda\lambda$3500-4700 Å at a resolution of 2.5 Å FWHM, and has been previously described in Rose (1985). An additional spectrum was obtained of NGC 205 by Lewis Jones at the KPNO coudé feed telescope in August, 1994 and kindly provided to us. This spectrum was recorded on a 800 x 800 pixel TI CCD at a resolution of 1.8 Å FWHM, and is the sum of three one-hour exposures. The wavelength region covered is $\lambda\lambda$3800-4400 Å. Second, a 20-minute integrated spectrum was obtained of the merger candidate NGC 3156 with a 800 x 1200 Loral CCD on the MMT in March, 1993 by J. Rose and Nelson Caldwell. This spectrum, which has a resolution of 7.2 Å, covers the $\lambda\lambda$3800-5400 wavelength region. Third, three one-hour spectra were obtained for the low-luminosity elliptical galaxy UGC 9519 with the KPNO 2.1-meter telescope by J. Rose and L. Jones in May, 1993. These spectra were recorded on a 3K x 1K Loral CCD at a spectral resolution of 2.1 Å and cover the wavelength region $\lambda\lambda$3450-4750 Å.

We also produced an "old galaxy" integrated spectrum from the composite spectra of 70 early-type galaxies in the central region of the Coma cluster. These spectra were obtained with the Hydra multi-fiber positioner, bench spectrograph, and Tektronics 2048 x 2048 CCD on the KPNO 4-m telescope, and have been described in detail in Caldwell et al. (1993). The spectra were recorded at a resolution of 4 Å and cover the wavelength region $\lambda\lambda$3500-5400 Å. The spectra used in the age-dating method are shown in Figure 2.

## 4. Determining the Ages of Starbursts

Starbursts are chiefly seen superimposed on an older stellar population. Thus to determine the age (and perhaps other parameters) of an observed starburst, it is necessary to accurately assess the contribution of the burst population to the total integrated light of the galaxy. In effect, then, one must determine the strength of the burst, where the burst strength depends on the star formation rate, duration, and initial mass function of the burst. While in principle this implies that the appearance of a starburst is a multi-parameter problem, we demonstrate below that a post-starburst spectrum depends primarily on the total amount of star formation in the burst and on the time elapsed since the end of the burst, and is relatively weakly dependent upon the details of the star formation history during the burst. As a result, the analysis of the E+A spectra in post-starburst galaxies is primarily a two parameter problem, if we ignore, for now, the effect of metal-abundance on the integrated spectrum of the burst.

### 4.1. Removing the Age-Strength Degeneracy

We begin by illustrating the fact that when model starburst spectra are combined with an old population spectrum in varying proportions to simulate bursts of different strengths, the degeneracy between age and burst strength mentioned in §1 becomes readily apparent. An example is shown in Figs. 3a and 3b, where the composite spectra for two combinations of post-starburst spectra and old population spectra are shown. The post-starburst spectra have been derived from the Bruzual/Charlot models as is described in §2.2 and the old population is the composite spectrum of 70 early-type galaxies in the Coma cluster described in §3. In both cases the length of the starburst is 0.3 Gyr. The post-starburst viewing time for the first spectrum is 0.0 Gyr and the spectrum is composed of 40% burst light and 60% old population light at 4000 Å. The second spectrum is viewed 1.5 Gyr after the burst ends and has an 80% burst light-20% old population light composition. The B - V colors for the two spectra are 0.729 and 0.767, respectively. Note how similar the shape of the Balmer lines are in the two spectra, which is further reflected in the fact that they have virtually the same value (0.774-0.779) for the H$\delta$/$\lambda$4045 index. Even though the two burst models are separated in age by 1.5 Gyr, they are very similar in terms of their colors and essentially indistinguishable in terms of their Balmer lines. On the other hand, note how different their Ca II line ratios are. This is further illustrated in Figure 3c, where the ratio of the two spectra has been plotted. In short, the Ca II index can be used to discriminate between the different burst situations, and in general to resolve the degeneracy between burst age and strength.

We have investigated the effect of burst age and strength on the behavior of the H$\delta$/$\lambda$4045 and Ca II indices in the following manner. We used as a starting point the 0.3 Gyr burst model spectra, at ages of 0.0, 0.5, 1.0, 1.5, and 2.0 Gyr after termination of the burst, that were described in §2.2. For each of these five post-starburst spectra, we linearly combined the burst model spectrum with the coadded Coma cluster early-type galaxy spectrum in a sequence covering from 100% burst light (and no old galaxy light) to 100% old galaxy light in 10% increments. Thus through these composite spectra, all possible burst strengths (defined as the fraction of light contributed by the burst at 4000 Å) have been investigated. The H$\delta$/$\lambda$4045 and Ca II indices were then calculated for each composite spectrum, thereby producing a trajectory of burst strengths for each post-starburst viewing time in the two dimensional spectral index space. In Fig. 4 are plotted the various trajectories of linear combinations in the Ca II versus H$\delta$/$\lambda$4045 diagram. The most important feature of the diagram is the evident decoupling of burst age from burst strength, i.e., from the position of a post-starburst galaxy in the Ca II versus H$\delta$/$\lambda$4045 diagram, both the age and strength of the starburst can be readily determined.

### 4.2. Age Determinations of Observed Galaxies

Given that with the spectral indices for a suspected post-starburst galaxy we can uniquely determine the age and strength of the starburst, we can test the age-dating method on the E+A galaxy spectra discussed in §3. We have plotted in Fig. 4 the observed indices for these galaxies. The $\pm 1\sigma$ error bars on the galaxy indices were obtained either from the scatter in indices found in multiple exposures of an individual galaxy (UGC 9519 and NGC 205) or by determining the S/N ratio in the sky-subtracted spectra (NGC 3156). While the three galaxies differ strongly in their Ca II and H$\delta$/$\lambda$4045 indices, all three lie in regions of the diagram that are covered by the model spectra. The one galaxy for which a good deal of independent data exists on its starburst status is NGC 205. In Fig. 4 both the UH 2.2-meter indices and the KPNO coudé feed indices place it between the 0.0 Gyr and 0.5 Gyr trajectories, indicating that the burst has just terminated. It is encouraging to see that the two NGC 205 spectra (obtained with different telescopes and detectors) produce approximately the same burst age. The fact that they imply different burst strengths relative to the underlying old galaxy can be attributed to the different effective aperture sizes used in the two observations. Our conclusion that a burst of star formation has recently concluded in NGC 205 is generally consistent with the fact that while OB stars are known to be present in NGC 205 (Baade 1951; Hodge 1973; Wilcots et al. 1990), indicating very recent star formation, the lack of emission lines in the optical spectra (Bica et al. 1990b) indicates that little gas remains there, hence the starburst must have ended. In fact, Gallagher and Mould (1981) have asserted that the age of the starburst is $\geq 2 \times 10^7$ yr. Their lower age limit is based on the absolute magnitude of the brightest blue stars and on the lack of M supergiants. On the other hand, a linear interpolation in Fig. 4 indicates a post-starburst age for NGC 205 of $\sim$0.1-0.3 Gyr. However, the evolution of the Ca II and H$\delta$/Fe I indices is undoubtedly far from linear during the first 0.5 Gyr, due to the steep dependence of main sequence lifetime on stellar mass at the high mass end of the IMF. Thus to the level of accuracy of the Gallagher and Mould (1981) lower age estimate and that of our still crude time resolution modelling of the Ca II index, the two age estimates are in basic agreement that star formation ceased very recently in NGC 205. Bica et al. (1990b) have performed a population synthesis of NGC 205 using a grid of star clusters spanning the age-metallicity plane. The resulting dominant population of the most probable solution is between $10^8$ and $5 \times 10^8$ yr in age with $[Z/Z_\odot]$ = -0.5. If they restrict their solution to one of solar metallicity for all but the oldest component (a less likely solution in their analysis), the dominant population turns out to be no older than $10^8$ yr old. While both age estimates are consistent with the one determined from the Ca II index, the solar metallicity restriction of the Bruzual/Charlot models prevents a detailed differentiation.

To further compare with other age determinations, we note that Schweizer and Seitzer (1992) have calculated a "heuristic merger age" for NGC 3156, based on the assumption that the galaxy is the product of a merger between two disk galaxies. Depending on the assumptions regarding the morphological type of the pre-merger galaxies and the efficiency of conversion of gas into stars at the time of the merger, they obtain merger ages in the range of 0.7 Gyr to 3.7 Gyr. Our post-starburst age of ∼0.8 Gyr for NGC 3156 lies within the estimated age range of Schweizer and Seitzer (1992). Moreover, Worthey (1995 and private communication) has determined a mean age of 1.0 Gyr, based on Balmer indices and the $C_2\lambda4668$ feature in Lick/IDS spectra. Again, the agreement is quite good. No independent data on the starburst nature of UGC 9519 is available.

### 4.3. Effects of Star Formation History on Age Determinations

In the above discussion it was implicitly assumed that the three observed E+A galaxies have truly undergone starbursts, i.e., we have not addressed the possibility that they are basically normal spiral galaxies whose star formation has been sharply truncated, perhaps by a gas stripping mechanism. To simulate the latter scenario, we have produced models in which the star formation rate is constant over 15 Gyr and then is suddenly terminated. This truncated 15 Gyr constant SFR population is then viewed at 0.0, 0.5, 1.0, 1.5, and 2.0 Gyr after the star formation is terminated. The same set of linear combinations of "burst" and old galaxy spectra have been made as in the case for the 0.3 Gyr burst models, and the Ca II and H$\delta$/$\lambda$4045 indices have been measured for each linear combination. The linear combinations thus simulate different proportions of truncated disk and underlying old bulge light. In Fig. 5 the indices for the observed E+A galaxies are plotted on the trajectories of the 15 Gyr constant SFR models. As can be seen in Fig. 5, if the Scalo (1986) IMF is used both NGC 205 and NGC 3156 lie outside the region of the index space that is covered by the models. Hence the possibility that they are truncated spirals is ruled out, and instead one can definitively conclude that starbursts took place in each of them. However, we also include in Fig. 5 the trajectories for a different IMF, namely the Salpeter (1955) IMF. With this IMF, NGC 205 is barely consistent with a truncated spiral. For both IMF's NGC 3156 is clearly a post-starburst system and UGC 9519 is consistent with a recently truncated spiral. While changing the IMF had virtually no effect on the 0.3 Gyr burst model results, over a longer burst such as the 15 Gyr one pictured in Fig. 5, the effect of changing the slope of the IMF at the low mass end becomes more pronounced.

¿From Fig. 5 it is apparent that while in some cases a starburst scenario can be readily distinguised from a truncated SFR scenario (e.g., NGC 3156), in other cases such a definite distinction is not possible (e.g., UGC 9519). It is clearly important to quantify under what circumstances such a distinction between a starburst and a truncated spiral can be definitively made. To do so, we have plotted in Fig. 6 the same 0.3 Gyr burst model trajectories as in Fig. 4. In addition we have plotted a thin solid line, labelled "CON", which represents the constant star formation models at various times after termination of star formation. The tick mark at the bottom of the line marks the index values at the moment of truncation of star formation, while each successive tick mark denotes time steps of 0.5 Gyr. Thus the line represents the expected evolutionary path in the diagram of a fading truncated spiral, under the assumption that there is no underlying old (i.e., bulge) population. If significant bulge light were present, then the integrated indices of the fading spiral would lie somewhere between the solid line and the old composite Coma apex point in the upper right of the diagram, depending on the balance between disk and bulge light. Hence any galaxies with indices to the left of the truncated spiral line (e.g., NGC 3156) cannot possibly be fading spirals and must instead have experienced a recent starburst. Since the fading truncated spiral line crosses the 0.3 Gyr burst trajectories at typically the 50% burst-50% old population level, we can thus conclude that post-starburst galaxy can be distinguished from a truncated SFR spiral if the burst is currently contributing at least 50% of the integrated light at 4000 Å.

A constant SFR with no underlying old population represents an extreme form of a truncated spiral. Caldwell et al. (1991) and Kennicutt et al. (1994) have demonstrated that while late-type spiral and irregular galaxies do indeed follow, on average, a constant SFR history, early-type spirals have undergone a substantially declining SFR over the Hubble time, with present-epoch SFR's typically 1%-10% of their original values. Consequently, we have also simulated an early-type spiral by running Bruzual & Charlot (1995) models with an exponentially declining SFR. We chose the e-folding time in the exponential so that the SFR at the truncation time (t = 15 Gyr) is 10% of the SFR at the beginning of the galaxy's lifetime (t = 0). The exponentially decreasing spiral models have been plotted as the thin solid line labelled "DEC" in Fig. 6. As can be seen in Fig. 6, the truncated declining SFR spiral can be readily distinguished from a post-starburst scenario in almost all cases, i.e., unless the burst contributes less than ∼25% of the integrated light at 4000 Å.

While Figs. 4, 5, and 6 demonstrate that it is possible to distinguish between a fairly strong starburst and a truncated spiral galaxy on the basis of the Ca II and H$\delta$/$\lambda$4045 indices, it is not yet clear to what extent the length of the starburst affects the ages and burst strengths derived from these indices. It is similarly unclear to what extent the detailed star formation history during the burst can affect the age determinations. To assess the effect of star formation history during the burst on the derived ages, we simply note that Newberry et al. (1990) have shown the B-V colors and H$\beta$ indices of their burst models to have very little dependence on whether the star formation rate during the burst is constant or Gaussian. To assess the impact of burst length on the derived ages, we recalculated all of the burst models in Fig. 4, but this time replaced the 0.3 Gyr burst with a 1.0 Gyr burst. The effect on the indices for the the two different starburst lengths can be seen in Fig. 7. Note the slight change toward the old, underlying population for the longer burst. However, the indices show a much greater sensitivity to the time elapsed since the burst ended rather than to the length of the burst. Thus the "starburst clock", for all intents and purposes, starts at the end of the burst, so that the age determinations made here set lower limits on the ages of the triggering events.

### 4.4. Effects of Metal Abundance on Age Determinations

Finally, a preliminary estimate can be made concerning the effect of metallicity on the age determinations presented here. For the burst populations we use as a starting point the theoretical isochrones for young stellar populations of Bertelli et al. (1994). We then assume that two burst populations with different metallicities but the same effective temperature for the main sequence turnoff will have the same Ca II index; thus we in effect assume that for young, hot turnoff populations, the effect of metal abundance on the stellar interior is far more important than that on atmospheric line blanketing. We then compare Bertelli et al. (1994) models for $Z = 0.02$ and $Z = 0.008$, which represents a difference in [Fe/H] of 0.4 dex, and find models with the same turnoff $T_{eff}$ at the two metallicities. For ages of $\sim$1-1.5 Gyr, there is about a 40% difference in age between the metal-poor and metal-rich models, in the sense that the metal-rich models are younger. In other words, our crude calculation suggests that if a starburst galaxy actually has [Fe/H]=-0.4, then we will underestimate the age of the starburst by $\sim$40% in applying the solar-metallicity Bruzual & Charlot (1995) models. This trend is supported by the population synthesis work of Bica et al. (1990b). As mentioned above in the discussion of NGC 205, the age of the dominant component of a model with $[Z/Z_\odot]$ = -0.5 versus one with solar metallicity decreased by $\sim$40-50%.

The metallicity of the underlying old population does not play a significant role in determining the ages of post-starburst galaxies. The calibration of the Ca II index performed in Rose (1984, 1985) which consisted of field dwarfs ranging in metallicity from [Fe/H]=-1.25 to [Fe/H]=+0.5, demonstrated that the Ca II index has a constant value for late-type stars *independent* of metallicity throughout the sample. Schmidt et al. (1995) investigated how different types of underlying populations affected the behavior of numerous spectral features in starburst scenarios. Specifically, they used an elliptical galaxy template with $[Z/Z\odot]$ = -0.5 and a combination of globular cluster spectra in the range -2$\leq$[Z/Z$\odot$]$\leq$-1. The behavior of the equivalent width of the Ca II K absorption line was essentially the same in both cases with a slight reduction in the dynamic range for the more metal poor underlying population. The starburst age and strength had a much more pronounced effect on the Ca II K line than the metallicity of the underlying population. Also Vigroux et al. (1989) measured the H$\delta/\lambda$4045 index as a function of internal velocity dispersion in a sample of galaxies. They found that over a range in dispersion from $\sim$100-300 km s$^{-1}$, the H$\delta/\lambda$4045 index changes by only 0.1. Thus even if this effect is entirely driven by metallicity, we can see that changing the [Fe/H] of an old galaxy population does not produce a large change in the H$\delta/\lambda$4045 index.

### 5. Summary

A method for determining the ages of starbursts in post-starburst galaxies has been developed which is based on the residual central intensities of neighboring spectral features, in particular, Ca II H + H$\epsilon$ and Ca II K. It allows an independent measure of both the burst strength and the burst age, which become degenerate when broadband colors or Balmer line equivalent widths are used as age discriminators. Three nearby post-starburst galaxies have been analyzed with this method and their ages are consistent with independent determinations, where available, and range from 0 to 1.5 Gyr. The results also show that at least 50% of the galaxy light at 4000 Å is from stars formed in the burst.

In a companion paper (Caldwell et al. 1995), the age-dating method is applied to post-starburst galaxies recently found in the Coma cluster. Since these Coma post-starburst galaxies are predominantly located in one region of the cluster, a common triggering mechanism is implied. Accurate ages for the starbursts can be compared with the dynamical timescale for the event which has been suggested by Burns et al. (1994).

The work carried out here would clearly benefit from further extension of the modelling techniques. Specifically, an exploration of non-solar abundance models would be highly desirable. Additionally, both the age and metallicity dependence of the Ca II index can be calibrated by acquiring integrated spectra for a number of young star clusters in the Galaxy and in the Magellanic Clouds (cf. Bica & Alloin 1986) with known ages and metallicities.

We wish to thank G. Bruzual for kindly supplying the Bruzual/Charlot (1995) models, without which this work would not have been possible and also for his patience and assistance in getting the models up and running at UNC. We also thank L. Jones for obtaining spectra of NGC 205 and of several hot stars with the KPNO coudé feed, and S. Charlot for useful discussions on the starburst modelling. This research was partially supported by NSF grant AST-9320723 to the University of North Carolina-Chapel Hill.

Fig. 1.— Comparison of a synthetic starburst spectrum with a stellar spectrum. (a) 0.3 Gyr starburst viewed 1.0 Gyr after completion of the burst. (b) spectrum of an A5 V star (HR6570) kindly provided by L. Jones. Note the prominent metal features evident in the burst spectra that are drastically reduced in the stellar spectrum. Identified spectral lines are as follows: L1 - Ca II K, L2 - Ca II H + H$\epsilon$, L3 - H$\delta$, L4 - G-band, L5 - H$\gamma$.

Fig. 2.— Plots of spectra used in age-dating method. (a) composite spectrum of 70 red Coma cluster galaxies, (b) UGC 9519, (c) NGC 3156, (d) NGC 205 (KPNO spectrum), (e) NGC 205 (UH spectrum)

Fig. 3.— Plots of (a) 0.3 Gyr starburst viewed immediately after burst completion with 40% of the light at 4000 Å coming from the burst and 60% of the light arising from an old population spectrum, (b) 0.5 Gyr burst viewed 1.5 Gyr after burst completion with 80% burst light and 20% old population light, and (c) the ratio of the preceding composite spectra.

Fig. 4.— The Ca II index is plotted versus the H$\delta$/$\lambda$4045 index for a 0.3 Gyr starburst length. The filled and open triangles refer to the MMT and KPNO data points for NGC 205, the open square refers to the spectrum of NGC3156, and the open pentagon denotes the data point for UGC 9519. The dotted trajectories denote linear combinations of the observed Coma cluster old galaxy spectrum (large filled circle) with model burst spectra 0.0, 0.5, 1.0, 1.5, and 2.0 Gyr after completion of a 0.3 Gyr burst (filled squares). The crosses represent 10% increments in the contribution of the burst spectrum to the integrated light at 4000 Å.

Fig. 5.— The Ca II index is plotted versus H$\delta$/$\lambda$4045 for a 15 Gyr burst length. The post-starburst galaxies are shown to be inconsistent with truncated constant SFR spiral galaxies. All symbols are same as in Fig. 4, except that solid lines refer to Salpeter IMF trajectories, while dotted lines, as in Fig. 4, are for Scalo IMF trajectories.

Fig. 6.— Same as Fig. 4 except plotted on top of the 0.3 Gyr burst trajectories are thin solid lines, marked "CON" and "DEC" which represent the evolutionary tracks of truncated spirals seen at different times after termination of star formation. The tick marks at the bottom point of the lines represents the index values right after the truncation of star formation, while each successive tick mark denotes a time step of 0.5 Gyr. The "CON" line represents the constant SFR scenario while the "DEC" line represents the exponentially declining SFR described in the text.

Fig. 7.— Comparison of index trajectories for a 0.3 Gyr burst (dotted lines) and a 1.0 Gyr burst (solid lines). Note the small movement toward an older population for the 1.0 Gyr burst.

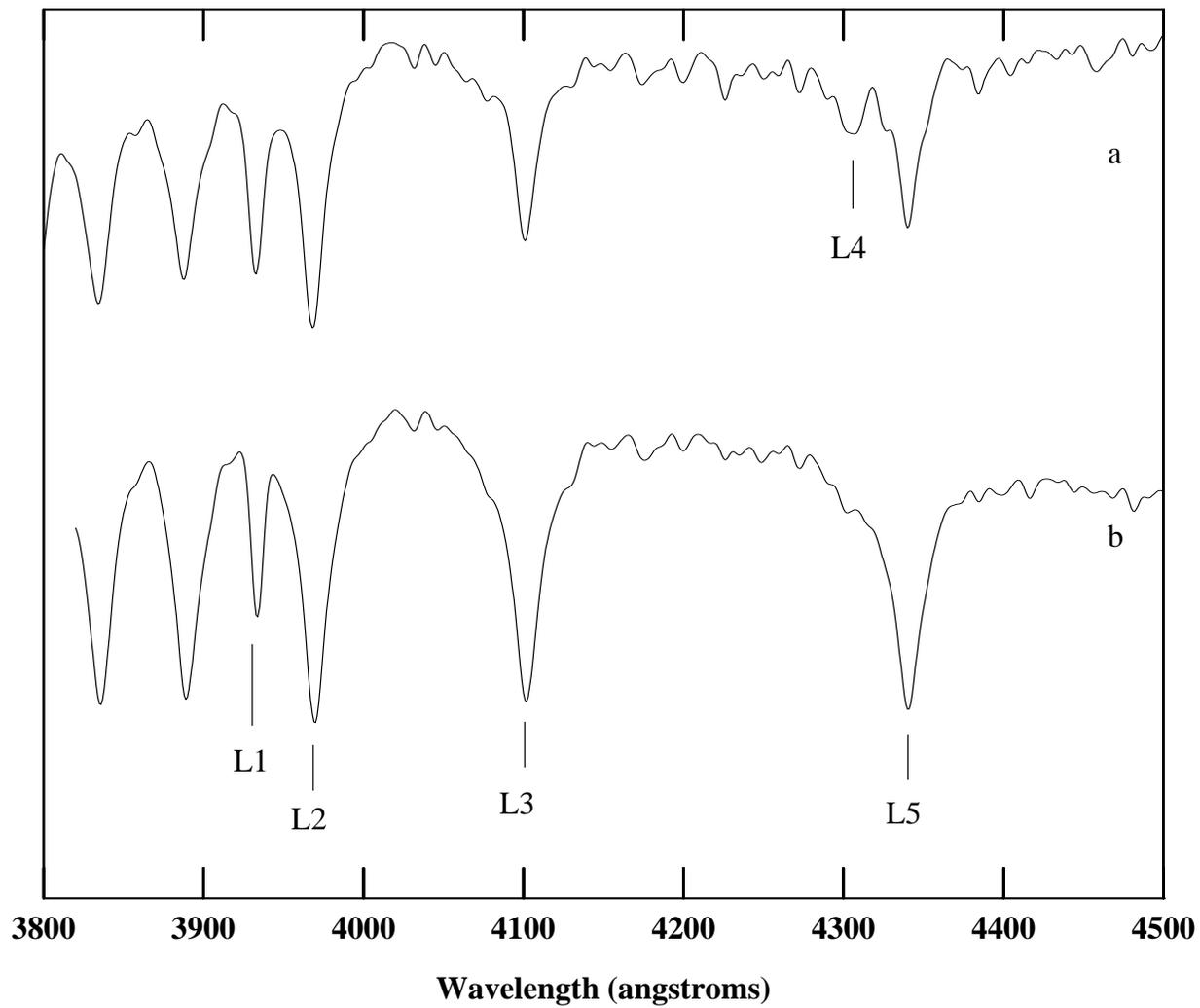

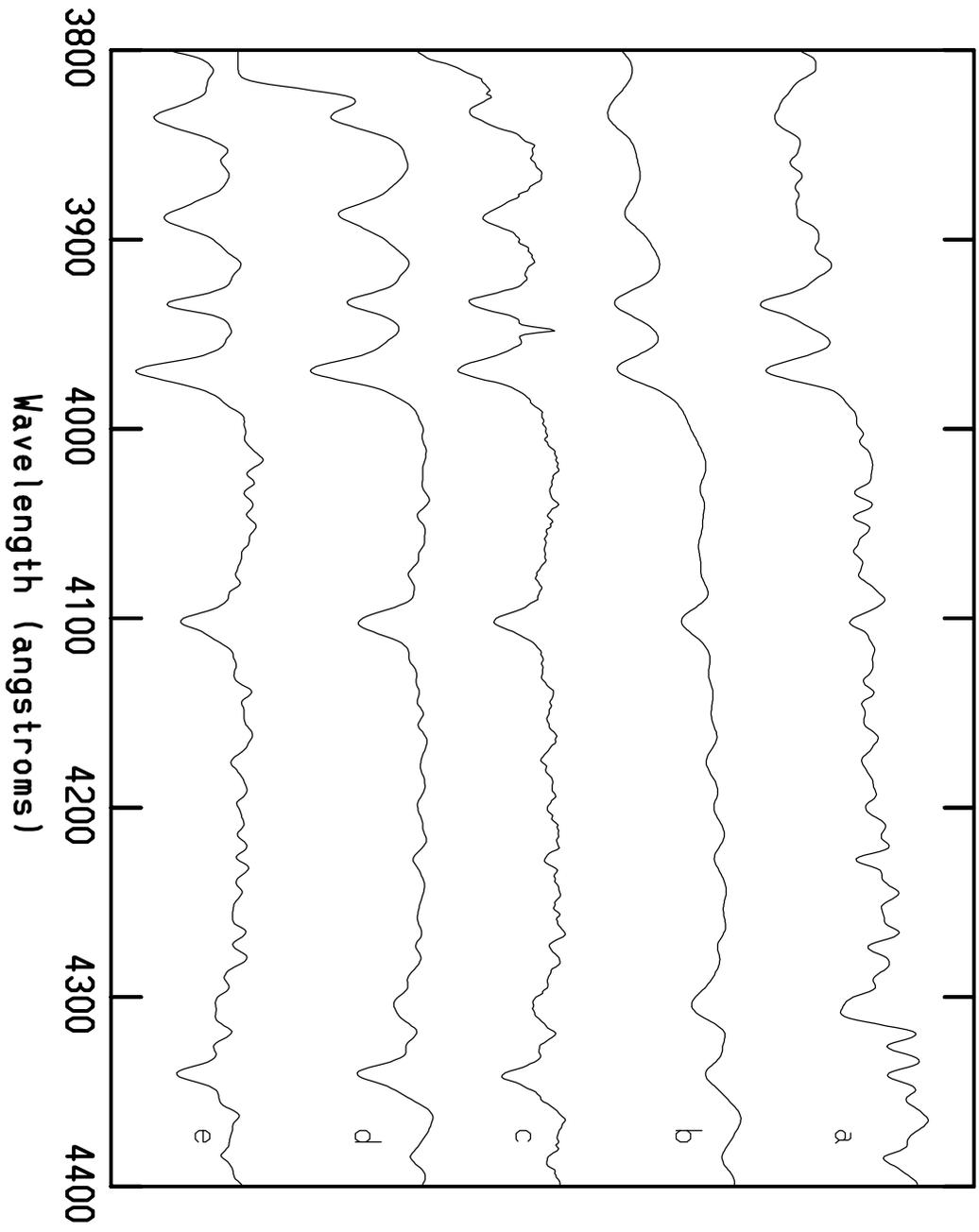

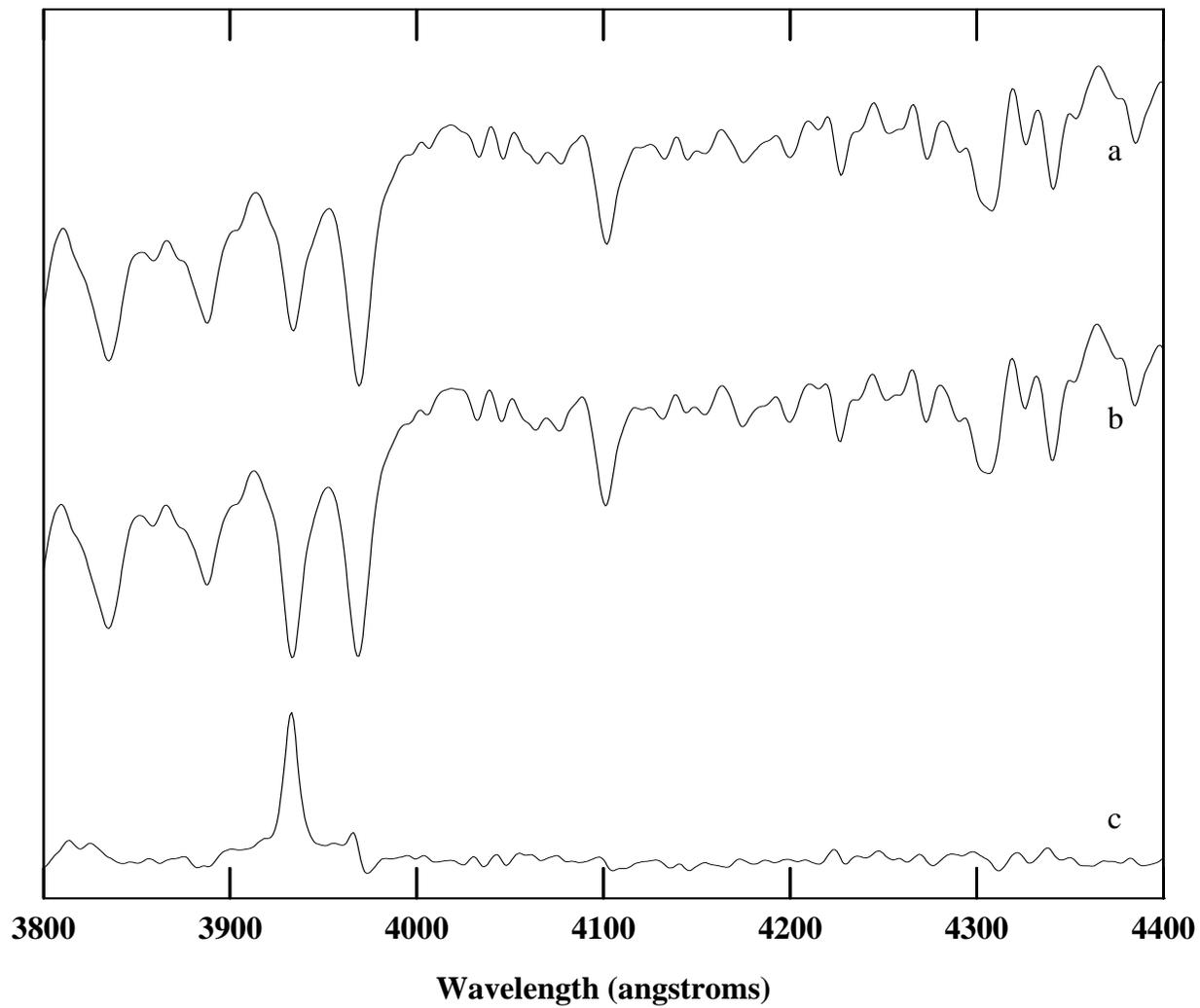

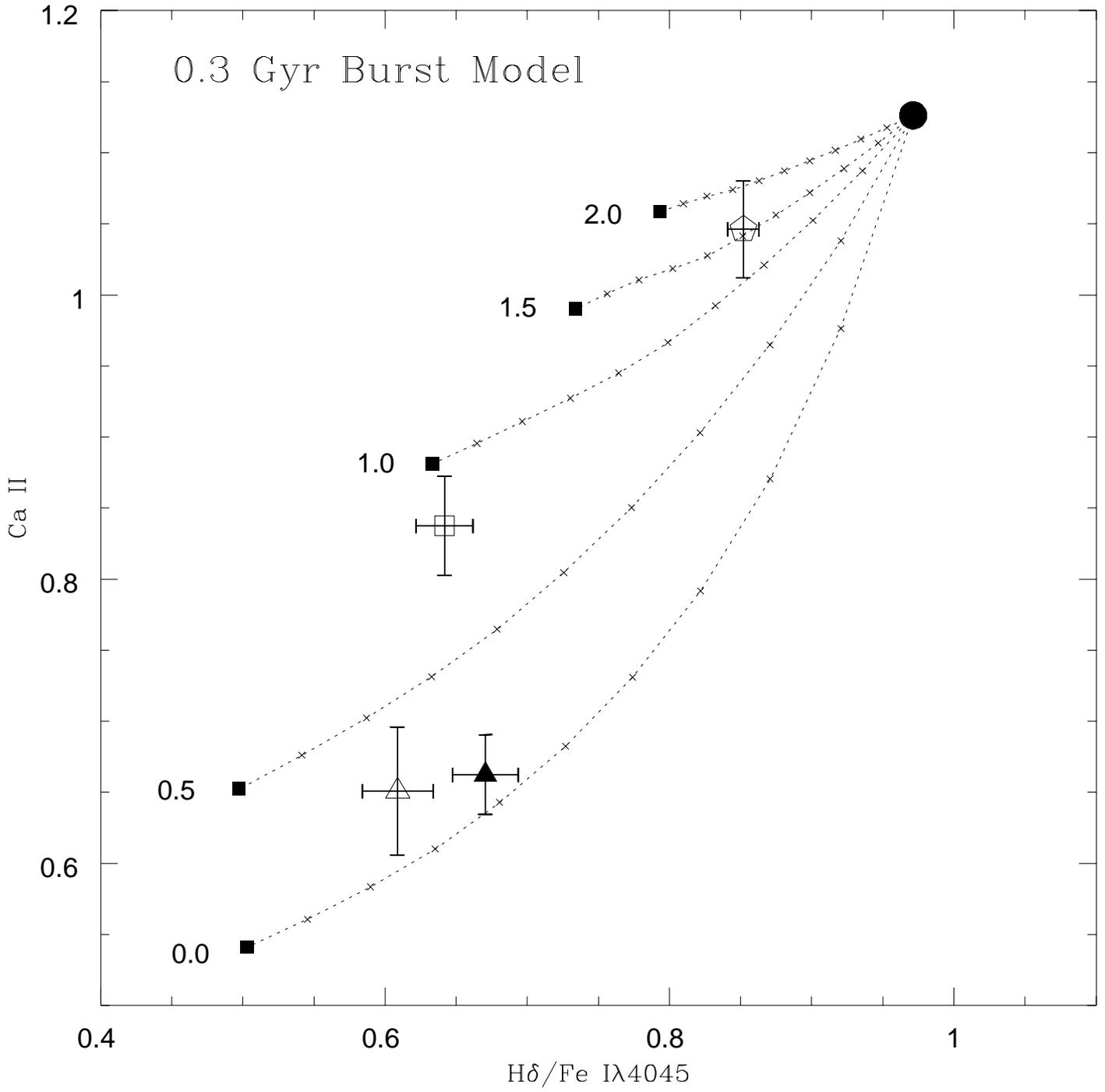

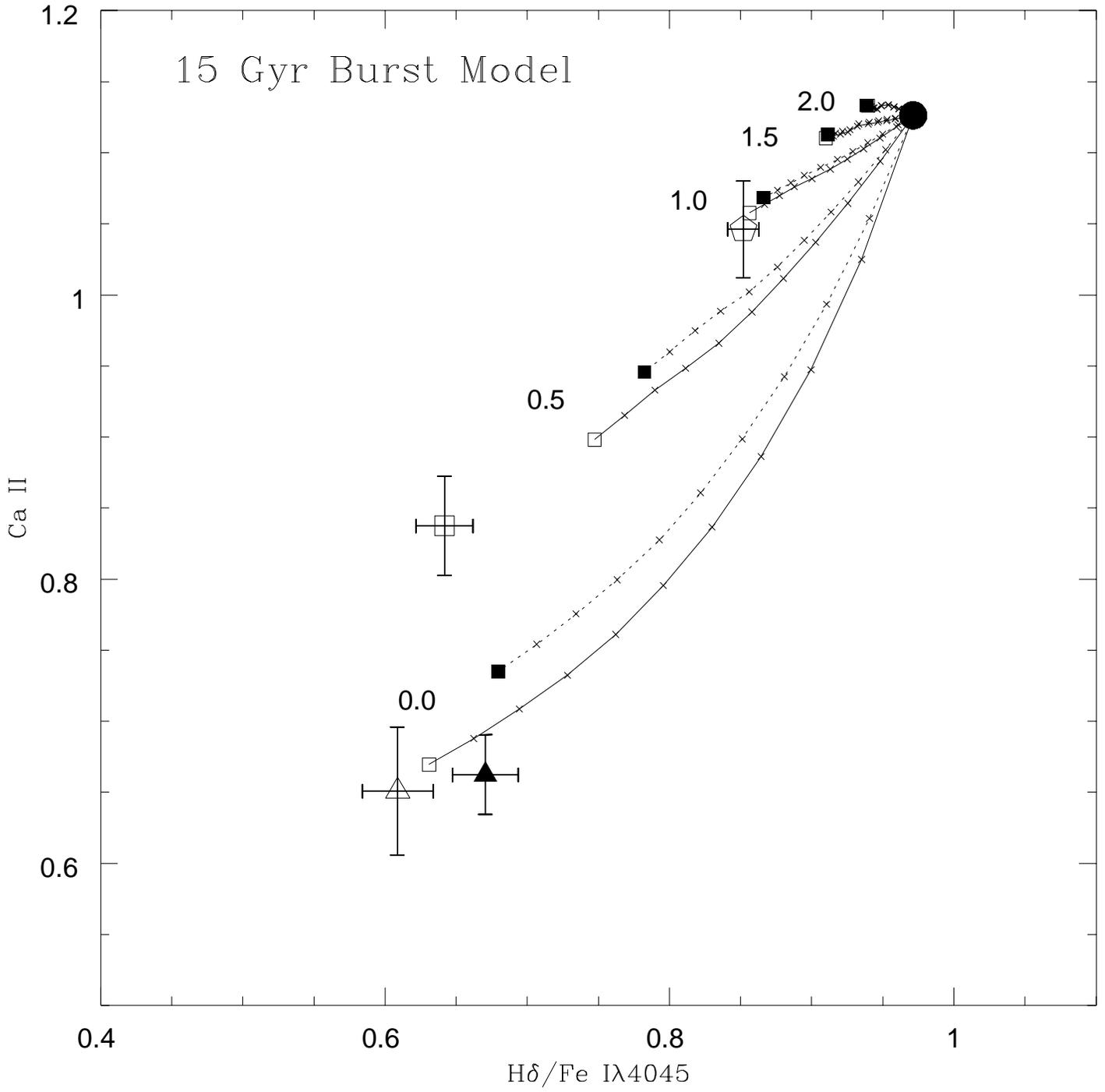

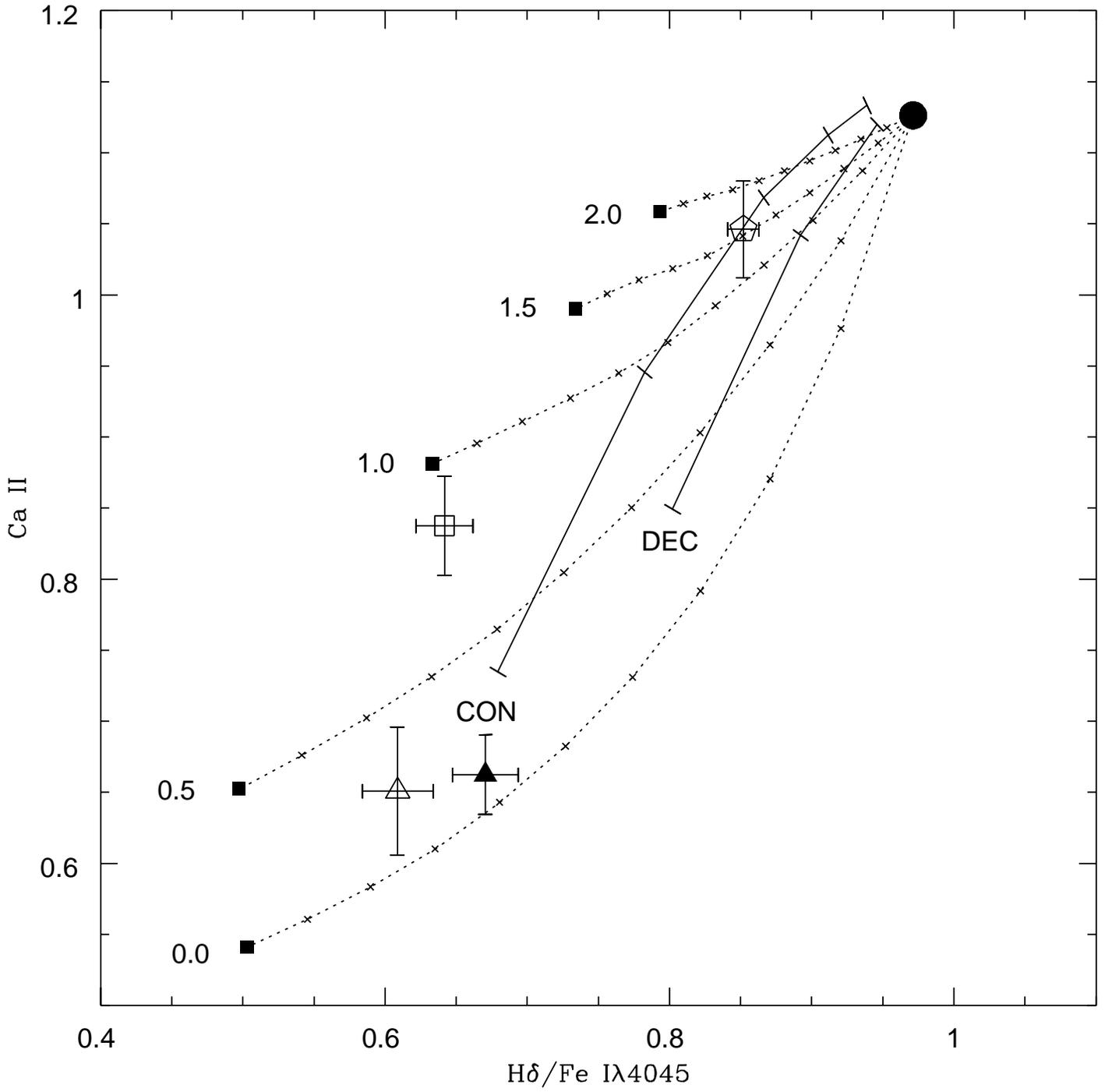

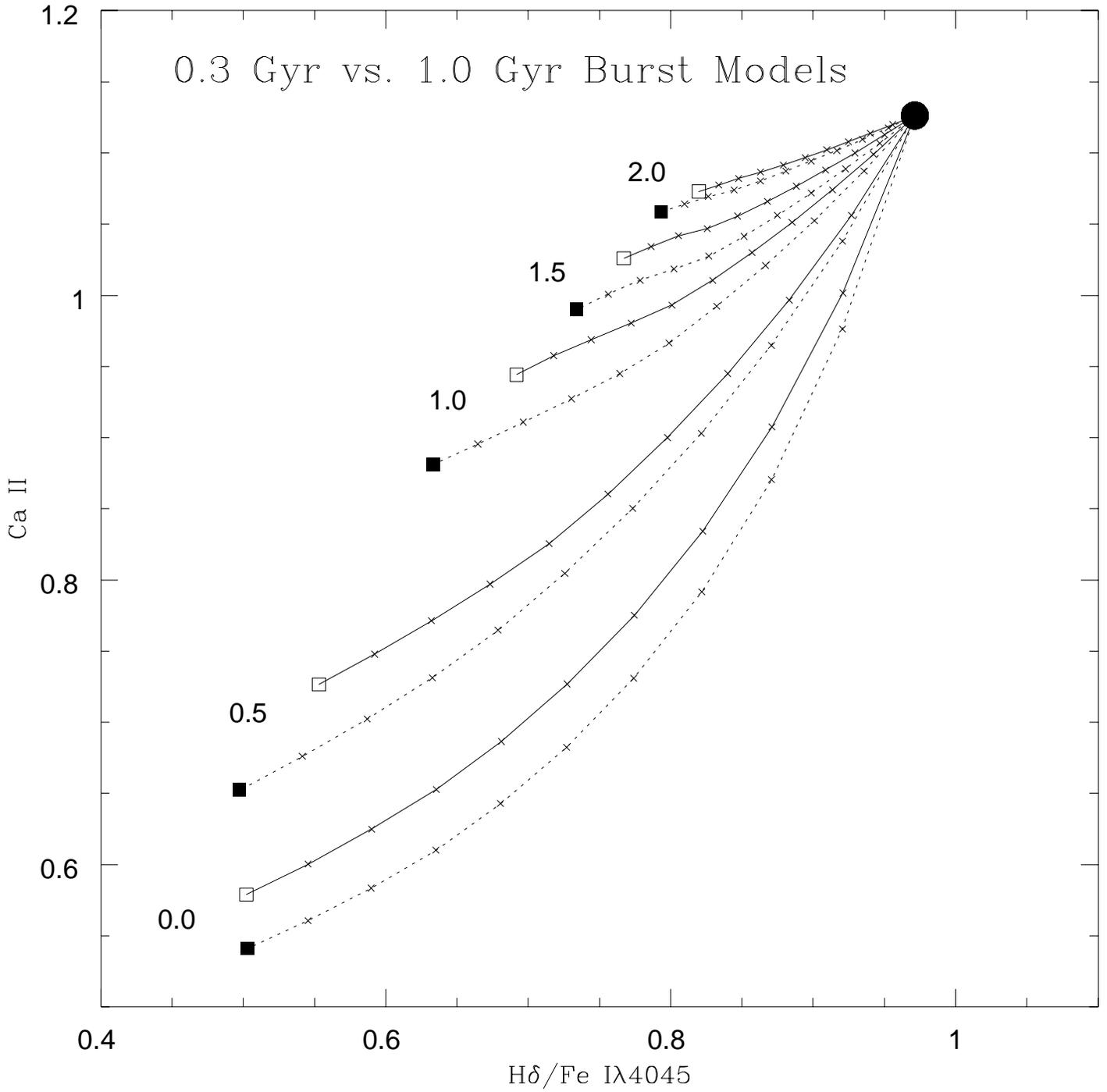